\documentclass{aa}

\usepackage{graphicx}
\usepackage{amssymb}

\usepackage{natbib}
\usepackage{url}

\bibpunct{(}{)}{;}{a}{}{,} 

\newcommand{\Msol}[0]{~$\mathrm{M}_{\odot}$}
\newcommand{\Msun}[0]{~$\mathrm{M}_{\odot}$}
\newcommand{\Sch}[0]{Schwarzschild{ }}

\newcommand{\eg}[0]{e.g.{ }}
\newcommand{\ie}[0]{i.e.{ }}

\newcommand{\eqn}[1]{\begin{equation} #1 \end{equation}}
\newcommand{\basel}{{\it B}a{\it S}e{\it L}}

\newcommand{\cmd}{CM-diagram~}
\newcommand{\cm}{\mathrm{CM}}

\newcommand{\hrd}{HR-diagram~}
\newcommand{\teff}{T_{\mathrm{eff}}}

\newcommand{\Hp}{\mathrm{H_{p}}}

\newcommand{\I}{\mathrm{I}}

\newcommand{\disp}{\displaystyle} 

\newcommand{\ec}[2]{^{#2}\mathrm{#1}}

\begin{document}
\title{The Faint Cepheids of the Small Magellanic Cloud:
                an evolutionary selection effect?}
\author{D. Cordier\inst{2,3}, MJ Goupil\inst{1} and Y. Lebreton\inst{2}.}
  
\offprints{daniel.cordier@ensc-rennes.fr}
  
\institute{LESIA,
           Observatoire de Paris-Meudon, 
           F-92195 Meudon Principal Cedex, France.
      \and GEPI,
           Observatoire de Paris-Meudon, 
           F-92195 Meudon Principal Cedex, France.
      \and \'Ecole Nationale Sup\'erieure de Chimie de Rennes, 
           Campus de Beaulieu, F-35700 Rennes, France.
          }

\date{Received ; accepted}

\markboth{Cordier et al., High magnitude SMC Cepheids}{}

\abstract{Two problems about the faintest Small Magellanic Cloud (SMC) Cepheids
are addressed. On one hand evolutionary tracks fail to cross the Cepheid
Instability Strip for the highest magnitudes (i.e. I-mag$\sim 17$) where Cepheids
are observed; Mass-Luminosity relations (ML) obtained from evolutionary tracks
disagree with Mass-Luminosity relations derived from observations. We find that
the above failures concern models built with standard input physics as well as
with non-standard ones. The present work suggests that towards highest
magnitudes, Cepheids stars undergo a selection effect caused by evolution: only
the most metal poor stars cross the Instability Strip during the ``blue loop''
phase and are therefore the only ones which can be observed at low luminosity.
This solution enables us to reproduce the shape of the lower part of the
Instability Strip and improves the agreement between observed and theoretical
ML-relations. Some issues are discussed, among them Beat Cepheids results
argue strongly in favor of our hypothesis.
\keywords{-- galaxies: SMC
          -- stars: evolution, variables}
}


\maketitle
  
\section{Introduction}

 Cepheids are variable stars located in the Color-Magnitude diagram 
(\cmd) within the Instability Strip (IS) where pulsation
phenomena take place via the $\kappa-$mechanism. Cepheid masses 
 approximately range between $\sim 3$ \Msun ~and $\sim 15$ \Msun.
 During the past decade, microlensing experiments as 
MACHO\footnote{\url{http://wwwmacho.mcmaster.ca}}, 
EROS\footnote{\url{http://www.lal.in2p3.fr/recherche/eros}}, 
MOA\footnote{\url{http://www.phys.vuw.ac.nz}} or
OGLE \footnote{\url{http://www.astrouw.edu.pl/~ogle/}} 
have produced a huge flow of data. As by-products of these 
observational programs, a large number of new variable stars, and among them,
Cepheids  have been detected. 
 OGLE 2 data  provide a large and high quality sample of Cepheids
belonging to the Small Magellanic Cloud (SMC) which we consider in this work. 
This extended and homogeneous data set has already put several shortcomings
of the SMC Cepheid modeling into light:

  (1) the evolutionary tracks built with standard input physics and for a
      chemical composition  $Z_{0}=0.004$ ($Y_{0}=0.251$) as usually assumed  for the SMC 
      fail to reproduce the observed Cepheid position within the \cmd for highest
      magnitudes (i.e. $\sim 17$ mag);

  (2) the Mass-Luminosity relation ($\mathrm{ML}^{\mathrm{puls}}$ hereafter)
      derived from pulsation properties \citep[see][]{beaulieu_etal_2001} and
      ML-relation from evolutionary tracks ($\mathrm{ML}^{\mathrm{evol}}$
      hereafter) do not agree.
\medskip 
 
The first problem arises because theoretical ``blue loops'' do not cross the
observed IS over the whole Cepheid mass range. Indeed an evolutionary track for
a mass about 5 \Msol~crosses observed IS three times, the first time (``first
crossing'') is the faster one -\eg $\sim 0.01$ Myr for an 5 \Msol~ model-, the
second time (``second crossing'') is the slower -\eg $\sim 0.20$ Myr- and the
third time  (``third crossing'') remains short, about $\sim 0.01$ Myr. These time
scale considerations tell us that the majority of the observed objects should be
in the second crossing stage. During this phase, the star burns He in its inner
regions. Third and second crossing both belong to the so-called ``blue loop''
excursion towards the blue side of Hertzsprung Russell (HR) diagram. Consequently
theoretical blue loops should cross the observed IS for the {\it entire} Cepheid
mass range. As we confirm in the first part of this work, theoretical tracks with
standard physics and free parameters varied in a reasonable range are not able to
provide blue loops which  reach the observed SMC Cepheids at low magnitude, i.e.
they are not able to model the low luminosity Cepheids in the SMC case.

 The second problem has been underlined -among others- by 
\citet{beaulieu_etal_2001} who found a strong disagreement between the
Mass-Luminosity relation $\mathrm{ML}^{\mathrm{puls}}$ and
$\mathrm{ML}^{\mathrm{evol}}$ in the SMC case. They determine a 
$\mathrm{ML}^{\mathrm{puls}}$ for the LMC and SMC
using  pulsation calculations, independently from  evolutionary calculations: for
a given Cepheid the mass $M_{\star}$ and the luminosity $L_{\star}$ are found
iteratively solving an equation of the type
$P_{i}^{\mathrm{theo}}(M_{\star},L_{\star})=P_{i}^{\mathrm{obs}}$,
where $P_{i}^{\mathrm{obs}}$ is the observed period ($i=0$ for fundamental
pulsators and $i=1$ first overtone ones) and $P_{i}^{\mathrm{theo}}$ the
theoretical one calculated with a pulsation code. The results of
\citet{beaulieu_etal_2001} are based on calculations that assume a metallicity
content $Z_{0}=0.004$ which is usually assumed to represent the mean metallicity of
the SMC.
 
 We are therefore led, in the second part of this paper, to propose  another
possibility and show that it can reconcile both issues: the blue loops at low mass
{\bf and} the Mass-Luminosity relation problem.  In our hypothesis
no high magnitude (i.e. with a mass around $\sim 3$ \Msol) SMC Cepheids
with a metallicity as high as the mean SMC value  can exist because  the
evolution  does not bring these stars far enough on the blue side to cross the
instability strip. The observed high magnitude SMC Cepheids must therefore be
undermetallic (i.e. $Z_{0} \sim 0.001$) with respect to the mean SMC metallicity
(i.e. $Z_{0} \sim 0.004$).

 In Sect.~\ref{stand_mod_obs}  we recall the  physical inputs used in our standard
models, which are similar to what is found in the recent literature. 
We next compare our evolutionary tracks with OGLE 2 observed Cepheids within a
$\cm$-diagram. Following a method similar to that used by
\citet{beaulieu_etal_2001}, we also compare $\mathrm{ML}^{\mathrm{evol}}$ and
$\mathrm{ML}^{\mathrm{puls}}$. In both cases we confirm the discrepancy.

 In Sect.~\ref{mod_uncert} we discuss the above issues in view of the
uncertainties of the standard models   and  discuss the effect of non-standard
physics in cases when models including such physics are available.
In Sect.~\ref{mod_z0001} we compare models calculated with $Z_{0} = 0.001$ with
observations.  Sect.~\ref{concl_discuss} is  devoted to discussion about the
possibility of the existence of SMC low luminosity Cepheids with metallicity as
low as $Z_{0} = 0.001$.


\begin{figure*}[t]
\rotatebox{-90}{\resizebox{8.8cm}{18cm}{
\includegraphics{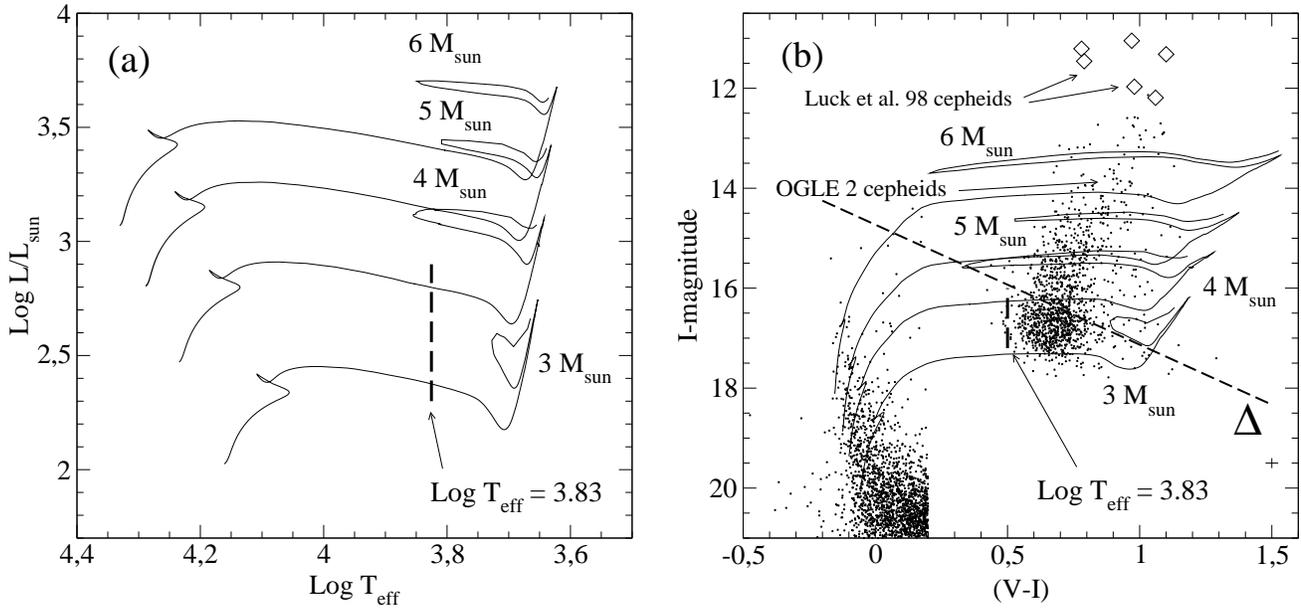}}}
\caption{\label{stand_theo_obs}(a) Theoretical HR  diagram showing our standard
evolutionary tracks with  masses 3.0, 4.0, 5.0 and 6.0 \Msol. 
The adopted  chemical composition is $X_{0}=0.745$ $Y_{0}=0.251$ $Z_{0}=0.004$. 
The heavy elements mixture composing $Z_{0}$ is from GN'93. The core overshooting
extension is 0.2 $\mathrm{H_{p}}$ along the entire tracks. (b) $\cm$-diagram
showing OGLE2 data (black dots) including the sample of Cepheids (fundamental and
first overtone pulsators). Only 10 \% of the stars belonging to the SC 1 OGLE 2
field main sequence are displayed for (V-I)$ < 0.2$.  The adopted reddening is
$\mathrm{E(V-I)=0.08}$ from \citet{udalski99}. The cross on the right side
indicates an estimation of errors: 30\% of $\mathrm{E(V-I)=0.08}$ on color and
0.1 mag on $\mathrm{I}$, for Cepheids. The effective temperature which should
be reached by models with masses of about 3 \Msol~ is indicated by a vertical dashed
straight line. The open diamonds are the sample of Cepheids used by
\citet{luck_et_al98} in their study of chemical composition, values for
$\mathrm{I}$-magnitude and $\mathrm{(V-I)}$ are those from \citet{luck_et_al98}'s
Table.~3 In panel (b), the  evolutionary tracks of (a) are also plotted
(solid lines). The straight dashed line $\Delta$ represents the limit below which
the evolutionary tracks fail to model the observed Cepheids. }
\end{figure*} 

\section{\label{stand_mod_obs}Standard models versus observations}

\subsection{Inputs for Standard evolutionary models}

 Our evolutionary models are built with the 1D Henyey type code 
CESAM\footnote{CESAM : Code d'Evolution Stellaire Adaptatif et Modulaire}
originally written by \citet{Mor97} in which we brought several improvements. 

 The equation of state is from \citet{Eggleton73} and the external boundary
condition is defined in a simplified model atmosphere involving the Eddington
$T(\tau)$ law.
The nuclear network involves 30 nuclear reactions, we have followed
\citet{schaller_et_al92} who used the same network as in \citet{maeder83}
for H-burning and \citet{maeder_meynet_87} He-burning network supplemented
with the $\mathrm{^{17}O(\alpha,n)^{20}Ne}$ reaction.
Nuclear reaction rates are from \citet{Caughlan88} excepted
$\ec{C}{12}(\alpha,\gamma)\ec{O}{16}$, $\ec{O}{17}(p,\gamma)\ec{F}{18}$ 
from \citet{Caughlan85} and  $\ec{O}{17}(p,\alpha)\ec{N}{14}$ from
\citet{landre90}. More recent nuclear rates do exist: NACRE by \citet{angulo_etal_99},
however adopted rate for $\ec{C}{12}(\alpha, \gamma)\ec{O}{16}$ is quite similar
to NACRE one (a factor of about two higher than \citet{Caughlan88} and about
80\% of \citet{Caughlan88} one.
 
The adopted  mean chemical composition for the SMC is taken as 
$X_{0}=0.745$, $Y_{0}=0.251$ and $Z_{0}=0.004$, corresponding to 
a metal to helium enrichment of $\Delta Y_{0}/\Delta Z_{0} = 2$ (see for instance
\citet{peimbert00}), to a primordial helium $Y_{P}= 0.243$ \citep{Izotov97} and to 
$\mathrm{[Fe/H]}= -0.68$ \citep{luck_et_al98}. Elemental abundances correspond to
the \citet{Grevesse93} (GN93) mixture consistent with OPAL96 calculations.

 Opacities are from \citet{iglesias96} (OPAL96) for high temperatures
($T\ge 10,000$ K) and  \citet{alex_fergu94} for cooler domains.
We stress that the central chemical composition during the He burning phase
differs strongly from GN93 (\eg 50 \% of $\ec{C}{12}$ and 50 \% of $\ec{O}{16}$).
Thus we have use opacity tables allowing a variable composition in $\ec{C}{12}$ and
$\ec{O}{16}$ with the aim of modeling the core as realistically as possible.
These tables have been built with \citet{magee95} elemental opacities (Los Alamos).

 The convective flux is computed   according to the prescription of  the
Mixing Length Theory \citep{bohm_vitense58}. The mixing length value
$l_{\mathrm{MLT}}$ -derived from solar calibration- is equal to 1.6  $\Hp$.
We used Schwarzschild's criterion to decide if the energy transport is radiative or
convective, and an extra mixing zone is added above the convective core (i.e. overshooting).
The extension of this zone is taken to be $l_{over}= 0.2$ $\Hp$ ($l_{over}= \alpha_{over}\;\Hp$).
 
 For the transformation of  theoretical quantities, ($M_{\mathrm bol}$,
$\teff$) into absolute magnitudes and  colors, we used
the  Basel Stellar Library ({\basel}, version  2.2)
of \citet{lejeune98} which provides color-calibrated theoretical
flux  distributions  for the  largest   possible range of  fundamental
stellar parameters, $T_{\rm eff}$ (2000 K to 50,000 K), $\log g$ (-1.0
to 5.5), and $\mathrm{[Fe/H]}$ (-5.0 to +1.0).\\

\subsection{Comparison between models and observations: the ``Blue Loop problem''}

 Cepheids data are from \citet{udalski99}. We have chosen to work with
$\mathrm{(V-I)}$ colors for which more data are available. Fundamental
and first overtone Cepheids are plotted in Fig.~\ref{stand_theo_obs}(b).
A mean $\mathrm{(V-I)}$ reddening is taken from \citet{udalski98a}, 
$\mathrm{E(V-I)= 0.08}$. The SMC distance modulus is fixed at $18.94$ from
\citet{laney_stobies94} with an internal error of $0.04\;\mathrm{mag}$,
this is a well accepted value, e.g. \citet{groenewegen_00b} found $19.11 \pm 0.11$
or $19.04 \pm 0.17$ depending on the photometric band.

In both figures (Fig.~\ref{stand_theo_obs}(a) and Fig.~\ref{stand_theo_obs}(b))
evolutionary tracks involving standard input physics are displayed. 
Fig~\ref{stand_theo_obs}(a) is a \hrd showing $\log L/L_{\sun}$ 
versus $\log \teff$. A segment of a line shows the temperature which
should be reached -according to OGLE observations-  by the evolutionary 
tracks for a stellar mass of about 3 \Msol; we will mention this mark in
further discussions.  An estimation of the uncertainties is also plotted 
in Fig.~\ref{stand_theo_obs}(b): we estimate the error on the colors to be  roughly
30\% of $\mathrm{E(V-I)}$ (which is the typical variation of reddening within
the sample) and assess an error of $0.1$ dex on $\I$-magnitude which roughly represents
the distance modulus uncertainty.  

 In Fig.~\ref{stand_theo_obs}(b) we have also plotted the stars from \citet{luck_et_al98}
for information.

 The general characteristics of these theoretical diagrams are similar to those shown by 
several groups like Geneva one, see \citet{charbonnel93}. This is not  
surprising because these authors have used similar physical inputs. For
instance,  the effective temperature at the tip of our 3 \Msol~ blue loop is
$\log \teff= 3.728$ which compares well with  \citet{charbonnel93} $\log \teff= 3.734$. 
In all cases, the $\log \teff$  value is far from the required one
of about $3.82$, \ie a temperature  hotter by $\sim 1200$ K.

 The main features put in evidence in Fig~\ref{stand_theo_obs} are: 

(1) the main sequence position seems to be reasonably well reproduced by the models

(2) the position of the blue tip of the 3 \Msol~ blue loop is too red. For an  $\I$-magnitude
corresponding to a mass of about 3 \Msol, we can clearly see a bulge of Cepheids. In fact
93\% of fundamental pulsators and 81\% of first overtone pulsators are located between
$\mbox{I}_{min} \sim 16.5$ and $\mbox{I}_{max} \sim 17.7$. Such a large  amount of objects
-statistically significant- cannot be explained solely by
$\sim 4.0$ \Msol~ first crossing models. Indeed, for a 4.0 \Msol~ standard model,
the time spent during the first crossing is $\tau_{first}= 4.6 \times 10^{-2}$
Myr while  the time it takes for the second and third crossing is 
$\tau_{\mbox{second crossing}}+\tau_{\mbox{third crossing}}= 3.433$ Myr.
 Hence, blue loops should cross the entire observational IS for the lowest masses.
The adopted value of distance modulus $\mu$ does not affect this conclusion. Indeed,
even if we take extreme evaluations: $\mu_{1} =18.66\pm 0.16 $ from \citet{udalski98a}
and $\mu_{2} = 19.05\pm 0.13$ from \citet{kovacs00}, the evolutionary track for $\sim$ 3
\Msol~ does not extend through   the observational IS.

\subsection{Comparison between observed and calculated Mass-Luminosity relations}

\begin{center}
\begin{figure}[t]
\resizebox{8.8cm}{11.38cm}{
\includegraphics{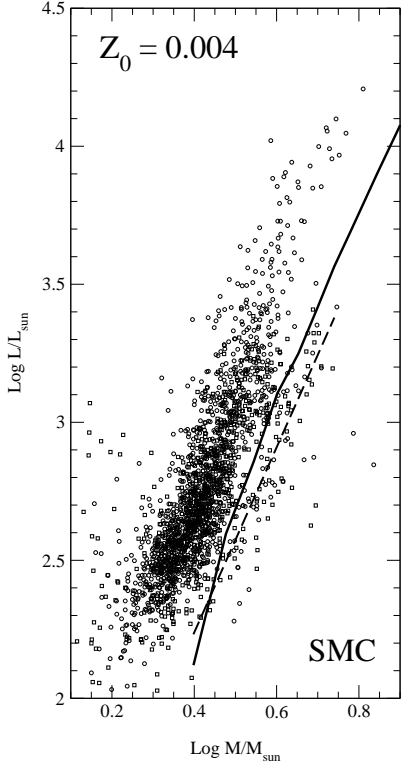}}
\caption{\label{rel_ML_z0004}Mass-Luminosity relation derived from OGLE observations
for $Z_{0}= 0.004$ corresponding to $[\mathrm{Fe/H}]\sim -0.7$. 
Circles: fundamental pulsators, squares: first overtone pulsators. 
Solid line: mass-luminosity relation from our evolutionary code with $Z_{0}= 0.004$
and an overshooting amount $\alpha_{\mathrm{over}}= 0.2 \;\Hp$; dashed line:
mass-luminosity relation from \citet{bono_et_al_00a}.}
\end{figure}
\end{center}
\subsubsection{Deriving Mass-Luminosity Relations from observations}

In order to derive a ML-relation 
from the observations, we use a method very similar to the one used by Beaulieu et
al. (2001). For each object we solve
iteratively the equation:
\begin{eqnarray}
   P_{i}^{\mathrm{theo}}( M_{\star}, 
          L_{\star}, T_{\mathrm{eff}}, Y_{0}, Z_{0}) = P_{i}^{\mathrm{obs}}
\end{eqnarray}
where $P_{i}^{\mathrm{obs}}$ is the observed period value ($i=0$ for fundamental
pulsators and $i=1$ for first overtones) and $P_{i}^{\mathrm{theo}}( M_{\star}, 
L_{\star}, T_{\mathrm{eff}}, Y_{0}, Z_{0})$ the theoretical one, computed with
the Florida LNA\footnote{Linear Non-Adiabatic} pulsation code which is a Castor
type code \citep[see][]{castor71}. During the iterative process, $M_{\star}$ is adjusted in
order to match $P_{i}^{\mathrm{obs}}$ and $P_{i}^{\mathrm{theo}}$, for a given iteration
$M_{\star}$ is fixed and we solve the following set of equations where the unknowns are
$(\log T_{\mathrm{eff}}, \log R_{\star}, \log L_{\star})$:
\begin{eqnarray}
\label{eq_teff}
\log T_{\mathrm{eff}} \;=\; 3.9224 \;+\; 0.0046 \log g \;+\; 0.0012 \; \mathrm{[Fe/H]}\\
  \nonumber -\; 0.2470 \;(V-I \;-\; (R_{V} \;-\; R_{I}) E(B-V)) 
\end{eqnarray}
\begin{eqnarray}
\label{eq_L}
2.5 \log L_{\star} = \mu_{\mathrm{SMC}} - V + R_{V}  E(B-V) + BC +
     4.75
\end{eqnarray}
\begin{eqnarray}
 L_{\star} = 4 \pi \sigma R_{\star}^{2} T_{\mathrm{eff}}^{4} 
\end{eqnarray}
\begin{eqnarray}
 g = G \frac{M_{\star}}{R_{\star}^{2}}
\end{eqnarray}
Eq.~\ref{eq_teff} comes from \citet{kovacs00} Eq.2, in which we brought absorption
corrections. Eq.~\ref{eq_L} is the Beaulieu et al.'s Eq.2. \citet{kovacs00} who made interpolations
of \citet{castelli_etal_97} stellar atmospheres models to convert magnitudes
into bolometric and effective temperature into colors. The luminosity $L_{\star}$ is in solar
units,
\begin{eqnarray}
BC = 0.0411 + 2.0727 \Delta T - 0.0274 \log g + 0.0482 \mathrm{[Fe/H]}\nonumber\\
      \nonumber - 8.0634 \Delta T^{2}
\end{eqnarray}
and $\Delta T = \log T_{\mathrm{eff}} \;-\; 3.772$. The magnitude and color $V$, $(V-I)$ are from
OGLE observations, $\mu_{\mathrm{SMC}}$ has been taken equal to $18.9$ consistently with
\citet{laney_stobies94}. Following \citet{udalski_etal_99_b} we took $E(B-V) \sim 0.08$,
$R_{V}= 3.24$ and $R_{I}= 1.96$.

 In order to apply this method one has to select the data. Indeed on the CCD detector, 
a Cepheid may be ``blended'' with another star, the magnitude of the object being
shifted towards lower magnitudes. These ``over-luminous'' objects lead to wrong
couples ($\log M_{\star}$,$\log L_{\star}$), therefore it is crucial to reject from
the sample the stars suspected to be blended with other object(s). From
OGLE data we have extracted amplitudes of pulsation in $B$, $V$ and $I$ bands
and then  derived $\mathrm{Amplitude}-\log P$ relations. The criterion to suspect
that an object is blended is the following: if a given object has a magnitude lower
than the mean magnitude (at least 0.2 mag lower) given by the $\mathrm{Magnitude}-\log P$ law
\textbf{and} an amplitude lower than the mean amplitude given by 
$\mathrm{Amplitude}-\log P$ relation,  this object is rejected.
Moreover we have also rejected some objects which appear to be suspiciously to red.
Finally we retain $1177$ fundamental pulsators and $709$ first overtone pulsators
and obtain similar samples of objects than \citet{beaulieu_etal_2001}.\\  

 Fig.~\ref{rel_ML_z0004} displays the resulting ML-relation derived 
with an assumed metallicity $Z_{0}=0.004$. We did not find significative
differences with results from \citet{beaulieu_etal_2001}.

\subsubsection{\label{uncert_ML}Uncertainties in Derived ML-relations}

 For years the question of Magellanic Clouds distance has been a subject of debate.
There were supporters for ``short'' distance scales -e.g. \citet{stanek_etal_98} with
$\mu_{\mathrm{LMC}} = 18.065 \pm 0.031 \pm 0.09$ mag- and for ``long'' distance scales
-e.g. \citet{laney_stobies94} with $\mu_{\mathrm{LMC}} = 18.53 \pm 0.04$ mag-
\citet{cioni_etal_00} derived a distance modulus for the LMC 
$\mu_{\mathrm{LMC}} = 18.53 \pm 0.04 \pm 0.08$ mag and \citet{mould_etal_00}
(HST Key Project Team) have adopted $\mu_{\mathrm{LMC}} = 18.50 \pm 0.04 \pm 0.15$ mag.
We made a test with ``short'' distance (i.e. $\mu_{\mathrm{SMC}} = 18.7$ mag); this leads
to differences in mass of $\delta \log M/M_{\odot} \sim -0.1$ and in luminosity
$\delta \log L/L_{\odot} \sim -0.08$ (consistently with Eq.~\ref{eq_L}).
We dismiss these ``short'' distance scales: (1) whatever is the technics used, recent
works seem in agreement with ``long'' distance scales; (2) a difference in mass
of $\delta \log M/M_{\odot} \sim -0.1$ would mean that evolutionary computations
would be completely wrong. We stress that \citet{beaulieu_etal_2001} have the same
point of view.
Therefore we adopted a ``long'' distance value for the SMC distance modulus:
$\mu_{\mathrm{SMC}} = 18.9 \pm 0.15$ mag; this choice is supported by the recent result of
\citet{harries_eta_03} who found $\mu_{\mathrm{SMC}} = 18.89 \pm 0.04$ (statistical)
$\pm 0.15$ (systematic) mag with a technics involving eclipsing binaries.

 Whereas the depth of the LMC seems to be negligible \citep{van_der_Marel_etal_01};
the depth of the SMC has been evaluated to range between $\sim 0.2$ and $\sim 0.4$
mag \citep{crowl_etal_01}. Then -for extreme cases- a given object inside the SMC
could have an actual distance modulus $+0.2$ mag larger or lower than
$\mu_{\mathrm{SMC}}= 18.9$ mag, which has to be regarded as an average value.
In order to estimate either the influence of an error on $\mu_{\mathrm{SMC}}$ or
an effect of SMC depth, we have made a test with $\mu_{\mathrm{SMC}}= 19.1$ mag we got 
$\delta \log M/M_{\odot} \sim +0.1$ and $\delta \log L/L_{\odot} \sim +0.08$.

 Another source of uncertainties is the reddening; if we assume an error
of $\pm 0.03$ mag on $\mathrm{E(B-V)}= 0.08$ mag, in turn we get a small uncertainty
on masses and luminosity: 
$\delta \log M/M_{\odot} \sim \pm 0.01$ and $\delta \log L/L_{\odot} \sim \pm 0.05$.

 Beside this, uncertainties connected to standardization of OGLE photometry are
clearly negligible; with $\delta V= \pm 0.02$ mag we obtained 
$\delta \log M/M_{\odot} \sim \pm 0.01$ and $\delta \log L/L_{\odot} \sim \mp 0.01$.

 Moreover \citet{beaulieu_etal_2001} made some additional tests: introducing turbulent
convection, computing non-linear models or changing the meshes size within models does
not yield to periods significantly different from those computed with LNA code.
Therefore the uncertainty on distance (error on $\mu_{\mathrm{SMC}}$ or effect of
SMC depth) remains the most important one.

\subsubsection{Comparison with ML-relations from evolutionary tracks}

 From Fig.~\ref{rel_ML_z0004}, we remark a large discrepancy between the ML-relations
derived from OGLE observations and from evolutionary calculations. For each
evolutionary track, luminosity has been read at the ``tip'' of the blue loop, locus
where the model spent at lot of time. The discrepancy is also found using
\citet{bono_et_al_00a}
mass-luminosity relation. The disagreement is getting worst when $\log M/M_{\odot}$
decreases. We must however emphasize that for $\log M/M_{\odot} \sim 0.4$
(i.e. $M/M_{\odot} \sim 2.5$), the evolutionary
track does not cross the Cepheid Instability Strip and a comparison between
$\mathrm{ML^{evol}}$ and $\mathrm{ML^{puls}}$ for $\log M/M_{\odot} \sim 0.4$
has no real meaning. Even an extreme value of $\mu_{\mathrm{SMC}}$ -i.e.
$19.1$ mag- can not lead to a perfect agreement between all ML-relations.

\section{\label{mod_uncert}Uncertainties in Standard Evolutionary Models}

 In this section, we review the factors affecting the blue loop extension. 
Before presenting any models, we briefly recall a method allowing some
predictions  about the blue loop extension. We
follow the work of \citet{lauterborn-et-al971a}b who have defined an ``effective
core potential'':
\eqn{\label{pot_eff}
     \Phi_{\mathrm{eff}} = \frac{\disp M_{c}}{\disp R_{c}} \; 
                  \mbox{e}^{(\alpha \; \Delta m \; \Delta X)}
    }
where $M_{c}$ and $R_{c}$ are respectively the mass and the radius of
the $\mathrm{H_{e}}$ core and $\alpha$ a constant. $\Delta m$ is the width of the zone
located between the $\mathrm{H_{e}}$ core and the beginning of the outer chemically
homogeneous region. $\Delta X$ represents the total hydrogen mass fraction variation
within $\Delta m$. Numerical experiments done by \citet{lauterborn-et-al971a}b have shown
that a  model undergoes a blue loop if this potential is lower than a critical value
$\Phi_{\mathrm{eff}}^{(crit)}$. We have to keep in mind this simple result: the lower
$\Phi_{\mathrm{eff}}$, the bluer the blue loop tip.
 
 In the next sections, we focus on a $3 M_\odot$ track because
the most severe discrepancy in the $\cm$-diagram is observed around this mass. 

\subsection{\label{overshoot}Overshooting}

 If we reduce the overshooting amount from $\alpha_{\mathrm{over}}= 0.2\;\mathrm{H_{p}}$
(``standard value'') to $\alpha_{\mathrm{over}}= 0.0\;\mathrm{H_{p}}$, the 
$\mathrm{H_{e}}$ core mass $M_{c}$ decreases as a consequence of the less extended H-core
on main sequence. As a consequence,  loops more extended toward the  blue are
expected. This is confirmed in Fig.~\ref{mosa}(b) where it clearly appears  that even
without any overshooting ($\alpha_{0}=0.0\;\mathrm{H_{p}}$), a 3 solar masses loop still
remains too short to account for the observational data.

\subsection{Mixing Length Parameter}

\begin{figure*}[h]
\rotatebox{0}{\resizebox{18cm}{22cm}{
\includegraphics{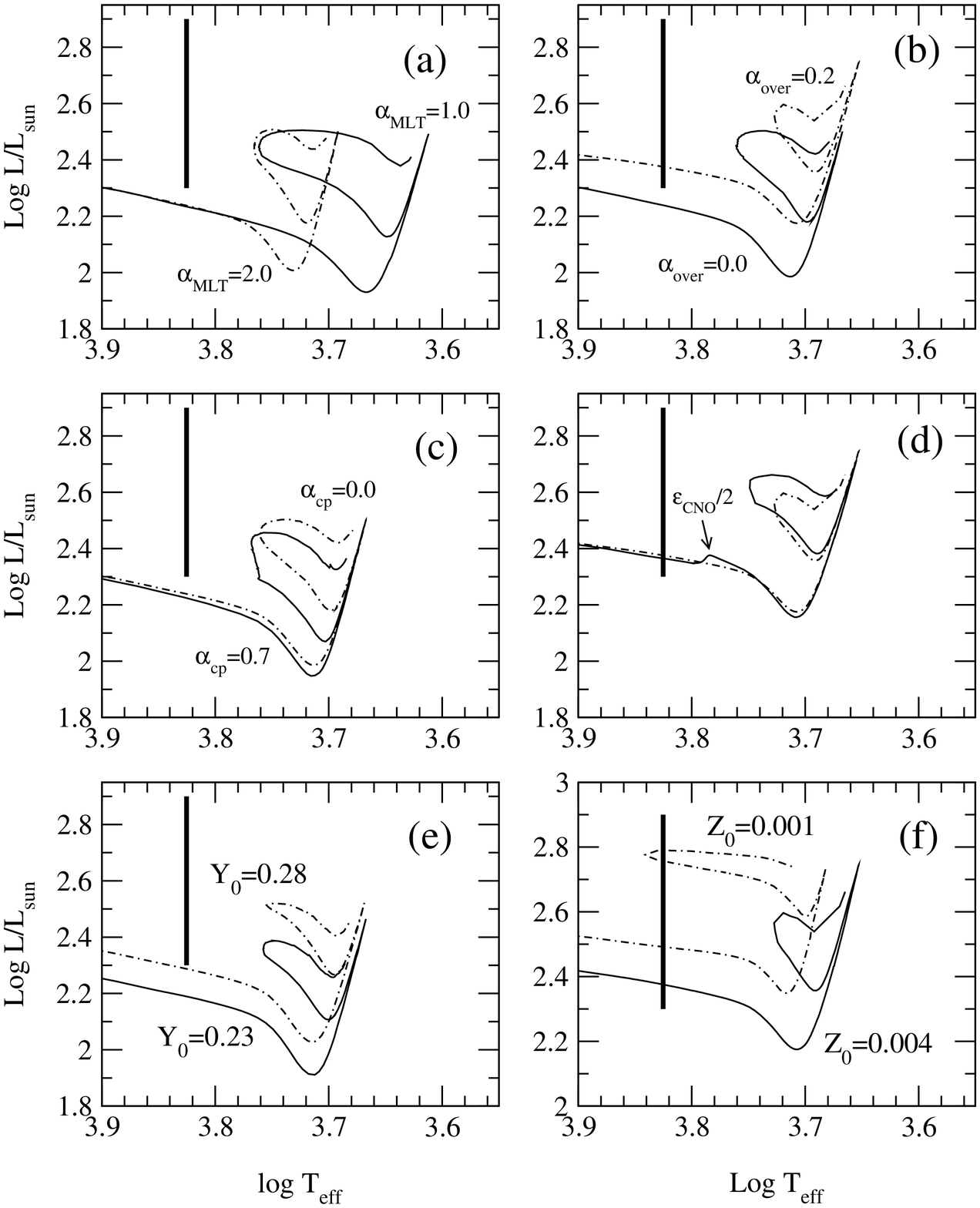}}}
\caption[]{\label{mosa}Influence of free  parameters on the blue loop
extension for a 3 \Msol model. The parameters are: 
(a) the mixing length, (b) the overshooting,
(c) the convective penetration (overshooting below external convective zone), (d) CNO
nuclear cycle energy production rate, (e) initial helium content and (f) initial
metallicity. For each plot, a vertical segment (defined in Fig.~\ref{stand_theo_obs}b)
shows the temperature which  the loop must reach  to cross over the entire
observational IS.}
\end{figure*}

The Mixing Length parameter $\alpha_{\mathrm{MLT}}=l_{\mathrm{MLT}}/\mathrm{H_{p}}$  has
been so far set equal to 1.6 in our standard models.
This value is derived from solar calibration (\citet{lebreton99}) and it is
probably not universal: it may depend on metallicity, mass, etc ... A priori,
$\alpha_{\mathrm{MLT}}$ acts only on the convective flux ($\alpha_{\mathrm{MLT}}$ is
involved in MLT temperature gradient calculation) and does not change the  position of
\Sch limit, hence $\Delta X$ in Eq.~\ref{pot_eff} should remain unchanged, more blueward
loops are not expected.  The tracks computed with the  extreme values
$\alpha_{\mathrm{MLT}}=1.0$ and $\alpha_{\mathrm{MLT}}=2.0$ are plotted in
Fig.~\ref{mosa}(a). Both tracks have been calculated with $\alpha_{\mathrm{over}}= 0.0$
which is the most favorable situation as explained in Sect.~\ref{overshoot}.
 As one can notice $\alpha_{\mathrm{MLT}}$ has a negligible influence on the blue tip
position. The effective temperature of the bluest point of the loop remains
approximatively
equal to $\sim 3.76$ (in Log) which is not enough to reach the warmer edge of
the observational IS, it still lacks $\sim 850$ K. We point out that a value
of $\alpha_{\mathrm{MLT}}=1.0$ is very unlike because it leads to a giant branch
around $\mathrm{(V-I)} \sim 1.5$ where there are no stars within $\cm$-diagram.


\subsection{Convective penetration}

 Similarly, although not identically as the overshooting process, turbulent eddies must
penetrate to some extent downward the convective envelope into stable radiative
regions. However, we do not know how far they penetrate. 

 Here we have carried out a calculation setting the extension of 
convective penetration at $\alpha_{\mathrm{cp}}= 0.7$ $\Hp$  following the
\citet{alongi_etal_91} prescription. They found that this value is needed to reproduce
the properties of red giants branch luminosity function. This amount (0.7 $\Hp$)
must be understood as an order of magnitude as \citet{alongi_etal_91}'s
calculations were performed before 1992 when OPAL group published his new opacity tables.

 During the giant branch (hereafter GB) episode, the convective penetration produces a
deeper penetration of the external convective zone. In this way, $\Delta X$ in
Eq.~\ref{pot_eff} decreases and  yields a lower $\Phi_{\mathrm{eff}}$ and bluer loop
tip. Evolutionary tracks are displayed in Fig.~\ref{mosa}(c) for a 3 \Msol~ without
overshooting (\ie both with $\alpha_{\mathrm{over}}= 0.0$ $\Hp$) and 
show -as expected- that convective penetration slightly extends  the loop but not 
enough to cross the entire IS. The extension difference reaches only a few $\sim 130$ K,
remaining too cold by $\sim 720$ K.
   \begin{figure}[t]
   \centering
\rotatebox{0}{\resizebox{8.8cm}{14.08cm}{\includegraphics{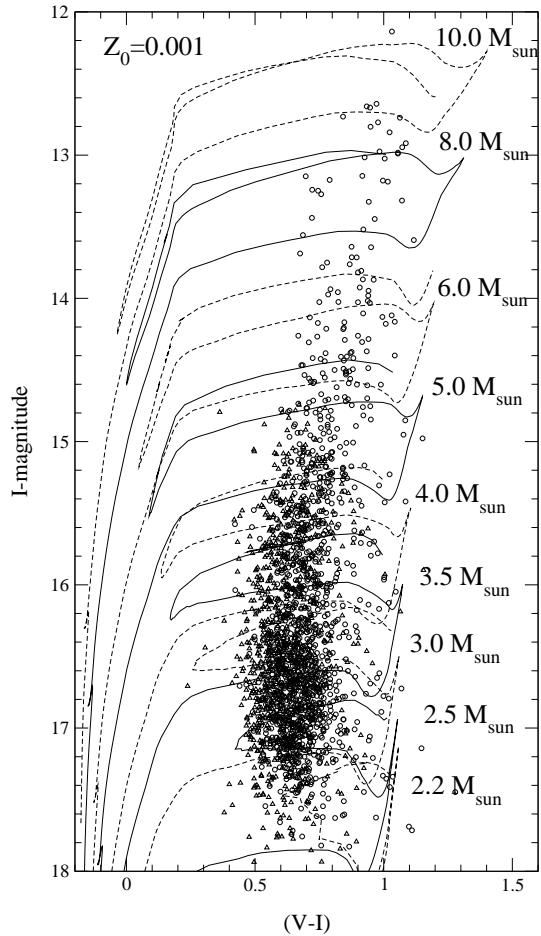}}}
      \caption{Grid of evolutionary tracks for $Z_{0}= 0.001$ with mass ranges between
               2.5 $\mathrm{M}_{\odot}$ and 10.0 $\mathrm{M}_{\odot}$. All the tracks
               cross the observational Instability Strip even the low mass tracks.
               Fundamental pulsators are represented by open circles and first
               overtone ones by open triangles.}
         \label{low_met_grid}
   \end{figure}
 
\subsection{Rotation}

\citet{maeder_meynet_01} present evolutionary tracks including
effect of stellar rotation at low metallicity $Z_{0}=0.004$ in the mass range 9.0 to 60.0
$\mathrm{M}_{\sun}$. The smaller mass value remains in the Cepheid domain.
In  Fig.6 of \citet{maeder_meynet_01} the reader can remark that the blue loop
extension is substantially reduced to $\log T_{\mathrm{eff}} \sim 4.12$ to 
$\log T_{\mathrm{eff}} \sim 3.95$ ($\Delta\log T_{\mathrm{eff}} \sim -0.17$) 
whereas we need $\Delta\log T_{\mathrm{eff}} \sim \textbf{+} 0.05$. This blue
loop reduction is due to the core extra-mixing added by convection equivalent
to an overshooting addition.

\subsection{\label{cno_infl}Influence of the CNO-cycle energy generation rate}

 Although the CNO nuclear reactions cycle is rather well known,
we have performed evolutionary calculations with an  energy generation rate 
$\epsilon_{\mathrm{cno}}$ artificially reduced by a factor two, from an arbitrary
chosen post-main sequence stage (indicated by an arrow in Fig.~\ref{mosa},d).
This magnitude of uncertainty (a factor of two) is extremely large because the
consequences during the Main Sequence phase would bring unavoidable disagreements 
between observations and models.

 One can again predict what can be expected from such  a numerical
experiment. A lower $\epsilon_{\mathrm{cno}}$ leads to a lower $M_{c}$ in
Eq.~\ref{pot_eff}, thus to a lower $\Phi_{\mathrm{eff}}$ and consequently to a bluer
loop. Fig.~\ref{mosa}(d) confirms this argument. Again, the loop is not extended
enough and even an unrealistic uncertainty of a factor of two on the global energy
generation rate $\epsilon_{\mathrm{cno}}$ cannot explain the disagreement between
observations and theory.

Enhancing by a factor two the $3\alpha$ reaction rate, 
is also found to have a negligible influence on the blue loop extension.

\subsection{\label{Y}Effects of helium}

 The initial helium content adopted, $Y_{0}$ is expected to have only a minor
 influence on the blue loop \textbf{extension}, indeed:
\begin{itemize}
 \item the central helium content $Y_{c}$ during the blue loop episode does not
       depend on the helium content of the initial homogeneous model $Y_{0}$. Hence
       $Y_{0}$ does not influence $\epsilon_{\mathrm{He}}$ the He-burning energy
       production rate because within the inner regions $Y_{c}=1-Z_{c}$ (with
       $Z_{c}$ the central heavy elements mass fraction) whatever the $Y_{0}$ value is.
 \item during the blue loop the H-burning shell moves through the
       ``$X$-profile'' where $Y$ varies between $\sim 1$ (boundary of
       He core) and $Y=Y_{0}$ (chemically homogeneous region mixed during
       the dredge-up episode when the model is closed to the Red Giants Branch).
       These Intermediate $Y$ values are independent of $Y_{0}$ (obviously excepted
       values closed to $Y_{0}$ itself)
\end{itemize}

 Therefore the influence  of $Y_{0}$ on the blue loop extension is expected to be very 
small. As a verification, models have been calculated with   $Y_{0}=0.23$ and
$Y_{0}=0.28$  which represent two extreme values: $Y_{0}=0.23$ is a rather low value
for primordial helium and  $Y_{0}=0.28$ which implies $\Delta Y_{0}/\Delta Z_{0}
\sim 9$ while ``reasonable'' values are around 2, for a review see \citet{luridiana_02}.
The tracks with $Y_{0}=0.23$ and $Y_{0}=0.28$ in (Fig.~\ref{mosa}(e))
show that the initial helium content has no influence on the blue loop extension:
the effective temperature  of the tip remains equal to 3.76, \ie $\sim 850$ K colder
than blue edge of IS, even in the favorable scheme of zero overshooting.

\subsection{\label{Z_effect}Effects of metallicity}

 The high sensivity of a blue loop extension to metallicity is well known. The physical
origin of this phenomenon is in the H-burning shell where material is processed
through CNO cycle. For a fixed heavy elements mixture (here GN93)  the
lower $Z_{0}$, the lower $X_{\mathrm{C}}$, $X_{\mathrm{N}}$, $X_{\mathrm{O}}$ 
(respectively C, N and O mass fractions) are. These three elements play the same role of 
catalysts in chemical reactions, therefore a C, N, O deficiency leads to a lower
energy generated. Then, $M_{c}$ in Eq.~\ref{pot_eff} remains lower for a longer time
and one obtains more extended blue loops. On one hand, the H-burning shell drives the
star structure on the Giant Branch, on the other hand, during the blue
loop episode, the $\mathrm{He}$-burning core pulls the model towards the blue edge,
where the $\mathrm{He}$ main sequence is located. The lower $\epsilon_{\mathrm{CNO}}$,
the stronger He central burning effect.

As a confirmation of this high metallicity sensivity, we have computed an evolution at
3 \Msol~taking a very low value: \ie $Z_{0}=0.001$ which corresponds to 
$\mathrm{[Fe/H]} \sim -1.3$. We compare the resulting extensions in
Fig.~\ref{mosa}(f). The blue loop crosses the entire IS, the tip reaching a 
position bluer than the blue edge of IS.

 Therefore the only way we have found to extend blue loops towards the high temperature
edge of the HR diagram is to decrease the metallicity. In the next section we compare
observational constraints and models built with $Z_{0}= 0.001$.

\section{\label{mod_z0001}Models with $Z_{0}= 0.001$}
\begin{center}
\begin{figure}[h]
\rotatebox{0}{\resizebox{8.8cm}{8.8cm}{\includegraphics{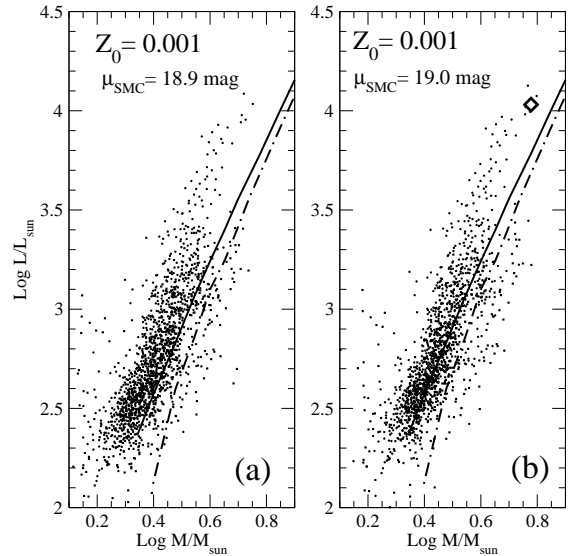}}}
\caption{\label{rel_ML_z0001}Mass-Luminosity relations derived from OGLE observations
assuming $Z_{0}= 0.001$ corresponding to $[\mathrm{Fe/H}]\sim -1.3$. Solid line:
mass-luminosity relation from our evolutionary code with $Z_{0}= 0.001$ and an
overshooting amount $\alpha_{\mathrm{over}}= 0.2 \mathrm{H_{p}}$; dot-dashed
line: $\mathrm{ML}^{\mathrm{evol}}$ for $Z_{0}= 0.004$ and  
$\alpha_{\mathrm{over}}= 0.2 \mathrm{H_{p}}$ (same as Fig.~\ref{rel_ML_z0004}, for comparison).
Panel (a): $\mathrm{ML}^{\mathrm{puls}}$ has been computed assuming a distance modulus
$\mu_{\mathrm{SMC}}= 18.9$ mag; panel (b): same thing assuming 
$\mu_{\mathrm{SMC}}= 19.0$. The diamond symbol shows the position of a 6 $\mathrm{M}_{\odot}$,
$Z_{0}= 0.001$ model with $\alpha_{\mathrm{over}}= 0.6 \;\mathrm{H_{p}}$.}
\end{figure}
\end{center}
\subsection{Blue loops at $Z_{0}= 0.001$}

 We have calculated a grid of evolutionary tracks at very low metallicity, i.e.
$Z_{0}= 0.001$. The results are displayed in Fig.~\ref{low_met_grid} where  one can
remark that the whole observed Instability Strip is crossed by the theoretical tracks,
even the fainter part, i.e. the lower region of the  color-magnitude diagram. These
results suggest that a great part of SMC Cepheids could be metal deficient compared
to the mean metallicity of the Small  Cloud.

 One interesting point is that the shape of the Instability Strip at high magnitude is
well reproduced by the decrease of the blue loop extension when going from
higher to lower masses.

\subsection{Mass-Luminosity relation at $Z_{0}= 0.001$}

 In Fig.~\ref{rel_ML_z0001} we have displayed the  Mass-Luminosity relations derived
from OGLE observations assuming a metallicity of  $Z_{0}= 0.001$. 
Fig.~\ref{rel_ML_z0001}b shows a better  agreement between $\mathrm{ML^{evol}}$ and
$\mathrm{ML^{puls}}$, if we assume $\mu_{\mathrm{SMC}} = 19.0$ mag -consistently
with recent determinations- the agreement for low masses is excellent. Unfortunately
it remains a discrepancy for higher masses -i.e. for $\log M/M_{\odot} \sim 0.7$,
this point will be discussed within the next section.
\begin{figure}[t]
\rotatebox{0}{\resizebox{8.8cm}{10.48cm}{
\includegraphics{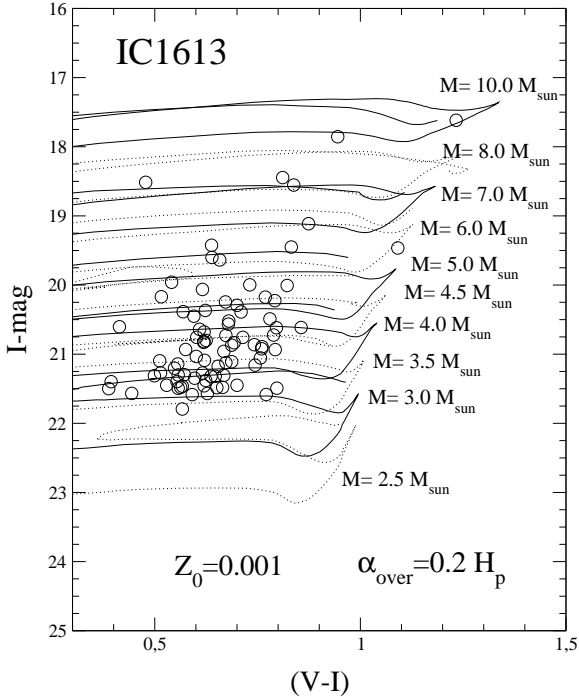}}}
\caption[]{\label{cmd_ic1613}Circles: Cepheids detected by OGLE team towards
IC1613, evolutionary tracks (solid and dotted lines) has been calculated with $Z_{0} = 0.001$ and
an overshooting amount $\alpha_{\mathrm{over}} = 0.2 \; \mathrm{H_{p}}$.}
\end{figure}

\section{\label{concl_discuss}Conclusion and discussion}

 Sect.~\ref{mod_z0001} shows that looking for an agreement between models and SMC
observations for both blue loop extensions and M-L relations for the SMC gives strong
hints that high magnitude (i.e. low mass) SMC Cepheids could be metal deficient
compared to the mean metallicity of the SMC; this fact could be explained
by a ``selection effect'': only stars with low enough metallicity could have an
evolutionary track crossing Cepheids Instability Strip. Unfortunately direct spectroscopic
determinations of $\mathrm{[Fe/H]}$ for the SMC Cepheids around I-mag $\sim 17$ are
not yet available. We discuss now a few issues in favor of the above proposition.
First we consider another metal poor galaxy and then discuss the
information which can be drawn from Cepheids in SMC clusters and beat Cepheids which
supports the existence of SMC Cepheids as metal poor as $\mathrm{[Fe/H]} \sim -1.3$ dex. 

\subsection{Comparison between our model and a very low metallicity Cepheid population}
\citet{udalski_etal_01} provided a sample of Cepheids belonging to the galaxy IC1613
with a metallicity $\mathrm{[Fe/H]} \sim -1.3\pm0.2$ \citep[see][]{lee_etal_93}. This
data set offers the opportunity to check whether our evolutionary models are valid for
metallicity as low as the value suspected for SMC Cepheids located below the $\Delta$
line in a CM-diagram. Fig.~\ref{cmd_ic1613} displays OGLE objects and our evolutionary
tracks for $Z_{0}=0.001$. The distance modulus for IC1613 is taken to be  $24.2\pm0.1$
mag and absorption is $\mathrm{A_{I}}= 0.05$ \citep[see][]{udalski_etal_01}; reddening
is given by \citet{schlegel_etal_98}. Fig.~\ref{cmd_ic1613} shows that the evolutionary
tracks cross the whole instability strip as defined by the observed Cepheids. For this
galaxy, no problem of blue loop exists with our models indicating that  the main features
of our models likely capture the essential evolutionary properties at low metallicity.
One can notice that even a cut-off of OGLE detector around I-mag $\sim 22$ can not really
change our conclusion because the blue loop for 2.5 $M_{\odot}$ is extended enough
to model Cepheids as faint as I-mag $\sim 22.2$.

\subsection{Evidences for metal poor stars within the SMC}

\citet{luck_et_al98} have determined the chemical composition
of Cepheids in SMC by means of high resolution spectroscopy. The
$[\mathrm{Fe/H}]$ values found by the authors range between $-0.84$ and $-0.65$
corresponding to a mass fraction $Z_{0}$ ranging between $\sim 0.0030$ and
$\sim 0.0045$, these values bracket the commonly assumed metallicity mean value for SMC,
i.e. $Z_{0}= 0.004$. However  two remarks are in order here: 
(1) the sample studied by \citet{luck_et_al98} has a quite poor statistic (6
objects) while OGLE sample contains about 2000 objects;
(2) more importantly, as shown by Fig.~\ref{stand_theo_obs}(b) the stars studied by
\citet{luck_et_al98} are objects much brighter than those around I-magn$\sim 17$
which are not reached by our blue loops. The reason for
the choice of bright objects is that spectroscopic determinations are easier
for brighter objects. Therefore any biais -concerning high magnitude SMC Cepheids
metallicity- cannot be excluded.

On the other hand, one may think that it is possible to infer some indirect information
about the SMC Cepheid metallicity. OGLE team has indeed discovered many stellar clusters
in the SMC, \citep[see][]{pietrzynski_etal_98}. Moreover \citet{pietrzynski_udalski_99}
have detected 132 Cepheids belonging to these clusters. One of the main properties of
stars belonging to a given cluster is to present the same chemical composition. Therefore
any indications about metallicity of these SMC clusters gives an information about the 
metallicity of Cepheids belonging to clusters. The literature is quite poor about 
metallicity determinations for SMC clusters. Table~\ref{det_smc_FesH} mainly taken
from \citet{crowl_etal_01} gives metallicity estimations for SMC clusters.
 Thanks to a cross identification we have found 2 clusters belonging to the catalogue
of \citet{pietrzynski_udalski_99} and having a metallicity determination in the
literature: NGC 330 (\verb+SMC0107+ in OGLE catalogue) and NGC 416 (\verb+SMC0158+).
However these clusters are known to be young or intermediate-age systems;
one has to check whether the ages of Cepheids are compatible with age of the 
cluster hosting them. 

 In the case of NGC 330 \citet{chiosi_etal_95} have found a maximum age of 48 Myr; beside
this \citet{pietrzynski_udalski_99_b} have derived from their study $31.6^{+8.2}_{-6.5}$ Myr.
From HST observations \citet{mighell_etal_98_b} estimate an absolute age for
NGC 416 of $6.6\pm0.5$ Gyr assuming that Lindsay 1 cluster is 9 Gyr old;
\citet{pietrzynski_udalski_99_b} confirm that NGC 416 is older than
1 Gyr. We underline that neither NGC 330 nor NGC 416 are mentioned in the erratum
\citet{pietrzynski_udalski_99_c}.

 For the Cepheid (\verb+SMC_SC7 206038+ in OGLE catalogue) suspected to belong to
NGC 330 if we assume a metallicity $\mathrm{[Fe/H] = -0.7}$ ($Z_{0}= 0.004$) and a mass
about 3.5 $M_{\odot}$ consistently with its position within CMD; we found an age
of $\sim 230$ Myr; assuming $\mathrm{[Fe/H] = -1.3}$ ($Z_{0}= 0.001$) and 3.0 $M_{\odot}$
we obtain $\sim 300$ Myr. Thus the Cepheid is likely a field star and does not
belong to NGC 330.

 Stars suspected to be NGC 416 objects have an I-magnitude between $15$ and $16$
mag; this corresponds to a mass around 4 $M_{\odot}$
for $\mathrm{[Fe/H] = -1.3}$ (metallicity consistent with cluster one). From our
evolutionary calculations we get -for such mass and metallicity- an age of $\sim 150$ Myr.

Unfortunately this value is not compatible with the estimated age for NGC416; indeed the
age spread (probably around $\sim 0.5$ Gyr) does not allow such young objects to belong
to the cluster. Nevertheless one can notice that the majority of SMC clusters in
Tab.~\ref{det_smc_FesH} are metal deficient, thus SMC stars with metallicity lower
than $Z_{0} \sim 0.004$ ($\mathrm{[Fe/H] \sim -0.7}$) does exist and the hypothesis
of metal deficient SMC Cepheids appears to be reasonable.

\begin{table}
\caption[]{\label{det_smc_FesH}Determinations of SMC clusters metallicity}
\begin{tabular}{lll}
\hline\noalign{\smallskip}
Cluster       &  $[\mathrm{Fe/H}]$         & Cepheid(s)?  \\
\hline\noalign{\smallskip}
\textbf{NGC 330}&$-0.82\pm0.11$ \tiny{(a)} & {\bf no}\\
NGC 411       &  $-0.68\pm0.07$ \tiny{(b)} & no          \\
NGC 152       &  $-0.94\pm0.15$ \tiny{(c)} & no          \\
Lindsay 113   &  $-1.24\pm0.11$ \tiny{(d)} & no          \\
Kron 3        &  $-1.16\pm0.09$ \tiny{(d)} & no          \\
NGC 339       &  $-1.50\pm0.14$ \tiny{(d)} & no          \\
\textbf{NGC 416}&$-1.44\pm0.12$ \tiny{(d)} & {\bf no}\\
NGC 361       &  $-1.45\pm0.11$ \tiny{(d)} & no          \\
Lindsay 1     &  $-1.35\pm0.08$ \tiny{(d)} & no          \\
NGC 121       &  $-1.71\pm0.10$ \tiny{(d)} & no          \\
Kron 28       &  $-1.20\pm0.13$ \tiny{(e)} & no          \\
Lindsay 38    &  $-1.65\pm0.12$ \tiny{(e)} & no          \\
Kron 44       &  $-1.10\pm0.11$ \tiny{(e)} & no          \\
\noalign{\smallskip}
\hline
\end{tabular}
\begin{tabular}{l}
{\tiny(a)\citet{hill_99}}; {\tiny(b)\citet{alves_sarajedini_99}};
{\tiny(c)\citet{crowl_etal_01}};\\ 
{\tiny(d)\citet{mighell_etal_98}};
{\tiny(e)\citet{piatti_etal_01}}
\end{tabular}
\end{table}

\subsection{Information brought by SMC Beat Cepheids}
\begin{figure}[h]
\rotatebox{0}{\resizebox{8.8cm}{8.8cm}{
\includegraphics{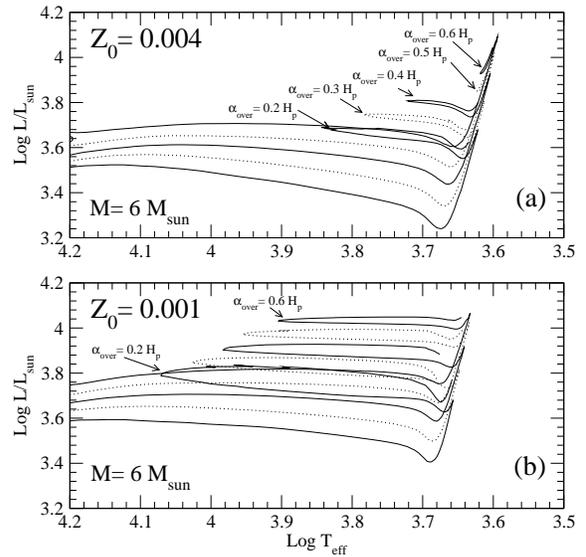}}}
\caption[]{\label{test_over_fig}Numerical experiments for a 6 $M_{\odot}$ model. For both
metallicities $Z_{0}=0.004$ (panel a) and $Z_{0}=0.001$ (panel b); we have computed a grid
of evolutionary tracks varying the overshooting amount from $0.2 \mathrm{H_{p}}$ up to
$0.6 \mathrm{H_{p}}$. For high metallicity blue loop disappears suddenly for overshooting
larger than $0.4 \mathrm{H_{p}}$ while the blue loop extension decreases monotonously in the
low metallicity case.}
\end{figure}

 OGLE Team has discovered a sample of 93 beat Cepheids in the SMC. 
\citet{udalski99} found that 23 pulsate simultaneously on the fundamental mode (hereafter F)
and the first overtone (hereafter 1OT), the remaining objects have been found to
pulsate simultaneously on the first and second overtones (hereafter 2OT). 

 In order to derive their $\mathrm{ML^{puls}}$ \citet{beaulieu_etal_2001} choose three
quantities among the four observational ones: $T_{\mathrm{eff}}$, $L$ and the periods
$P_{k}$ and $P_{k+1}$ ($P_{0}$ for F/1OT and $P_{1}$ for 1OT/2OT); they calculate the
theoretical value of $P_{k+1}$ which is noted $P_{k+1}(\mathrm{calc)}$ (the observed
one being $P_{k+1}(\mathrm{obs})$). They next define the parameter
$\epsilon = P_{k+1}(\mathrm{calc)}/P_{k+1}(\mathrm{obs})$) allowing a comparison
between theory and observations. They explore the influence of different
important parameters, particularly distance modulus and reddening and finally conclude
that a solution (i.e. $\epsilon \sim 1$) is found simultaneously for F/1OT pulsators
\textbf{and} 1OT/2OT \textbf{only} {\bf  if} the metallicity is settled to
$Z_{0} = 0.001$ (i.e.  $[\mathrm{Fe/H}] \sim -1.3$).

\begin{figure}[t]
\rotatebox{-90}{\resizebox{7cm}{8.8cm}{
\includegraphics{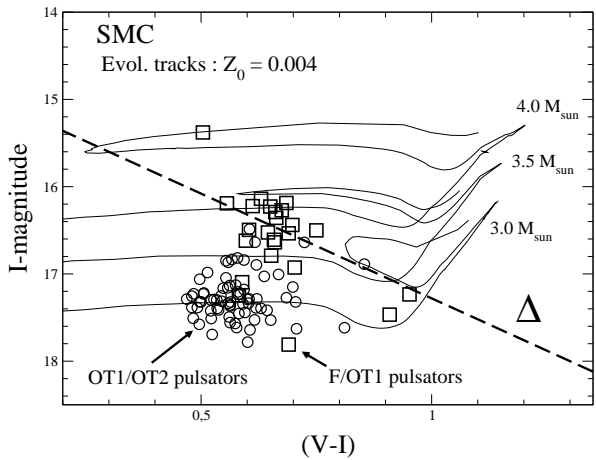}}}
\caption[]{\label{beat_ceph_fig}Location of SMC beat Cepheids detected
by OGLE team in the CM diagram. Squares are F/1OT pulsators and circles are 1OT/2OT
pulsators. The evolutionary tracks have been computed for the metallicity value
$Z_{0}= 0.004$ and the dashed line $\Delta$ limits the regions where the Cepheids are
expected to be metal-deficient ($Z_{0}= 0.001$).}
\end{figure}

 In Fig.~\ref{beat_ceph_fig} we have plotted the observed  beat Cepheids
together with evolutionary tracks and the straight line $\Delta$ defined in
Sect.~\ref{stand_mod_obs}. This plot clearly shows that the beat Cepheids are located
in the region where we suspect that objects are metal deficient (i.e. with a metallicity
around $Z_{0} \sim 0.001$). All 1OT/2OT pulsators are below $\Delta$ (excepted one
object) while F/1OT pulsators are scattered slightly above and below.

 Thus these pulsation/evolution models of SMC beat Cepheids argue in favor of a
relation between a metal deficiency (with respect to the mean value of the SMC) and the
existence of SMC Cepheids at low magnitude.

\subsection{The case of high mass Cepheid}

 Although it is slightly out of scope of this paper where we focus on faint SMC Cepheids, we
will debate in this paragraph the case of brighter objects: i.e. $\log M/M_{\odot} \sim 0.7- \sim0.8$
corresponding to $3.6 \lesssim \log L/L_{\odot} \lesssim 4.1$. Whatever is the assumed metallicity:
$Z_{0} = 0.004$ or $Z_{0} = 0.001$, few objects with an evaluated mass around $\log M/M_{\odot} \sim 0.77$
($M \sim 6 M_{\odot}$) have a $\mathrm{ML^{puls}}$ in discrepancy with the related 
$\mathrm{ML^{evol}}$.

 We can make some hypothesis: as shown in Sect.~\ref{uncert_ML} with a larger distance modulus
-i.e. $\mu_{\mathrm{SMC}}= 19.1$ mag- we get $\delta \log M/M_{\odot} \sim +0.1$ and
$\delta \log M/M_{\odot} \sim +0.08$ (compared with the situation with $18.9$ mag), but this
extreme value is not able to bring a full agreement between $\mathrm{ML^{puls}}$ and
$\mathrm{ML^{evol}}$ for brighter objects. On the other hand it is unlike that all these stars
would be located deeper in SMC than others.

 For a given mass value, a way to enhance the luminosity is to consider a larger overshooting amount.
As suggested by \citet{cordier_etal_02}, it can not be excluded that average overshooting amount
for intermediate mass stars increases when metallicity decreases; they have derived -assuming
$Z_{0}= 0.004$ for SMC main sequence stars- $\alpha_{\mathrm{over}}= 0.40^{+0.12}_{-0.06} \;\mathrm{H_{p}}$.
With a \textbf{LMC} bump Cepheids study \citet{keller_wood_02} infer an overshooting amount
$\Lambda_{\mathrm{c}}= 0.63 \pm 0.03 \;\mathrm{H_{p}}$ ($\sim 0.3 \;\mathrm{H_{p}}$ in our formalism).

 We made a test -involving all mass values- with $\alpha_{\mathrm{over}}= 0.4 \;\mathrm{H_{p}}$,
as expected blue loop extensions are
reduced (for low masses tracks the excursion of blue loop within IS is less deep) and luminosity
is increased not enough ($\delta \log L/L_{\odot} \sim +0.1$) to get an agreement between
$\mathrm{ML^{evol}}$ and $\mathrm{ML^{puls}}$ for $\log M/M_{\odot} \sim 0.77$. Then, to increase
the overshooting amount over the whole range of mass is not the solution.

 Another possibility is that overshooting can depend on mass, increasing with mass; as suggested for
instance by \citet{young_etal_01}. Thus we have concentrated us on $6 M_{\odot}$ models, varying
overshooting amount between $0.2 \;\mathrm{H_{p}}$ (our ``standard'' value here) up to
$0.6 \;\mathrm{H_{p}}$; this for both metallicities: $Z_{0} = 0.001$ and $Z_{0} = 0.004$.
Results are shown in Fig.~\ref{test_over_fig}; it is clear that beyond $0.4 \;\mathrm{H_{p}}$ one
no longer gets blue loop for $Z_{0} = 0.004$. In contrast with $Z_{0} = 0.001$ blue loop
extension decreases, but in a way where Cepheids IS is crossed by tracks. $\log L/L_{\odot}$
increases reaching large enough value (for $0.6 \;\mathrm{H_{p}}$, see diamond symbol in
Fig.~\ref{rel_ML_z0001}(b))
in order to make it up $\mathrm{ML^{evol}}$ and $\mathrm{ML^{puls}}$ for $\log M/M_{\odot} \sim 0.77$.
We stress that \citet{keller_wood_02} found their quite ``high'' overshooting amount
using a sample of bright LMC Cepheids, this support our proposal of an higher overshooting
for brighter Cepheids.
One more time our work favors a solution involving low metallicity. Towards high masses,
another selection effect could occur if the overshooting increases with mass and reached
$\sim 0.6 \;\mathrm{H_{p}}$ for masses larger than $\sim 6 M_{\odot}$ at low metallicity.
A detailed study is needed on this topic and is out of the main goal of this paper.
 
\medskip 

\subsection{Summary}
 In this paper we have explored two main problems related to the SMC Cepheid population:
(1) the blue loop extension for high magnitude stars, (2) the Mass-Luminosity relation.
We have first shown that blue loop extension is extended enough only if the metallicity
is substantially lower than the commonly used value for SMC objects models.

 Evolutionary tracks computed with $Z_{0}= 0.001$ correctly reproduce  the
Instability Strip shape for low masses and  Mass-Luminosity relation derived
from these tracks is in rather good agreement with $\mathrm{ML^{puls}}$ deduced from
observations using a technique similar to \citet{beaulieu_etal_2001} one.
 The remaining discrepancy for the small population of brighter objects could be
explained by a joined effect of low metallicity and rather enhanced core mixing
process. Further researches are needed on this subject. We emphasize that 
\citet{pietrukowicz_02} -who estimates period change rates of SMC OGLE Cepheids-
found also for brighter objects a rather bad agreement between models and observations.
 Our point is that all Cepheids below $\Delta$ are likely metal poor and stars
above $\Delta$ belong probably to a ``mixed'' population. Finally, the present work
strongly suggests the existence of an evolutionary selection effect for
fainter Cepheids belonging to SMC. High resolution spectroscopic chemical composition
determinations for SMC Cepheids through the entire IS and particularly around magnitude $17$
are requested to bring definitive arguments in favor or against the present suggestion.
This could be possible for a sample of few stars with UVES VLT spectrograph. Results 
would put a new light on the cosmologically important metallicity dependence of the
Cepheids Period-Magnitude relation.
\begin{acknowledgements}
We warmly thank the anonymous referee who contribute to the improvement of this paper
with his/her remarks and suggestions. We also thank Jean-Philippe Beaulieu and
J.Robert Buchler for valuable discussions; we are also grateful to the OGLE group for
providing their data and to people involved in \verb+GNU/Linux+ projects. D. Cordier
thanks E.N.S.C.R. for working facilities and his ``Am\'enagement de service''
during 2002$/$2003 Academic Year.
\end{acknowledgements}

\bibliographystyle{apj} 
\bibliography{apj-jour-perso,bibliographie_MASTER}

\end{document}